\documentclass[11pt]{article}
\usepackage{graphicx}
\usepackage{amsmath}
\usepackage{amsfonts}
\usepackage{amssymb}
\usepackage{epsfig}
\usepackage{times}
\usepackage{cite}
\usepackage{calc}
\usepackage{version}
\usepackage[english]{babel}
\usepackage{epsf}
\usepackage{graphics}
\usepackage{caption}
\usepackage{upgreek}
\usepackage{ulem}
\usepackage[dvips]{color}

\newcommand{\lqcd}{\Lambda_{_{\rm QCD}}}

\begin{document}

%%%%%%%%%%%%%%%%%%%%%%%%%%%%%%%%%%%%%%%%%%%%%%%%%%%%%%%%%%%%%%%%%%%%%%%%%%%%%%
%%%%%%%%%%%%%%%%%%%%%%%%%%%%%%%%%%%%%%%%%%%%%%%%%%%%%%%%%%%%%%%%%%%%%%%%%%%%%%
\begin{titlepage}

\null

\vskip 1.5cm
%\begin{flushright}
%Report: IFIC/11-18, FTUV-11-0419\\
%\end{flushright}

\vskip 1.cm

{\bf\large\baselineskip 20pt
\begin{center}
\begin{Large}
A Monte Carlo study of jet fragmentation functions in PbPb and pp collisions at $\boldsymbol{\sqrt{s}=2.76}$ TeV
\end{Large}
\end{center}
}
\vskip 1cm

\begin{center}
Redamy P\'erez-Ramos\footnote{
Department of Physics, P.O. Box 35, FI-40014 University of Jyv\"askyl\"a, Jyv\"askyl\"a, Finland}
\footnote{
Sorbonne Universit\'e, UPMC Univ Paris 06, UMR 7589, LPTHE, F-75005, Paris, France}
\footnote{
CNRS, UMR 7589, LPTHE, F-75005, Paris, France}
\footnote{
Postal address: LPTHE tour 13-14, $4^{\text{\`eme}}$ \'etage, UPMC Univ Paris 06, BP 126, 4 place Jussieu, F-75252 Paris Cedex 05 (France)}
\footnote{
e-mail: redamy.r.perez-ramos@jyu.fi, perez@lpthe.jussieu.fr},\quad Thorsten Renk
\footnote{
Department of Physics, P.O. Box 35, FI-40014 University of Jyv\"askyl\"a, Jyv\"askyl\"a, Finland}
\footnote{
Helsinki Institute of Physics, P.O. Box 64, FI-00014 University of Helsinki, Helsinki, Finland}
\footnote{
e-mail: thorsten.i.renk@jyu.fi, trenk@phy.duke.edu}
\end{center}

\baselineskip=15pt

\vskip 3.5cm

The parton-to-hadron fragmentation functions (FFs) obtained from the YAJEM and PYTHIA6 
Monte Carlo event generators, are studied for jets produced in a strongly-interacting 
medium and in the QCD ``vacuum" respectively. The medium modifications are studied with the {\sc YaJEM} code in 
two different scenarios by (i) accounting for the medium induced virtuality 
$\Delta Q^2$ transferred to the leading parton from the medium, and (ii) 
by altering the infrared sector in the Borghini-Wiedemann approach. The results of our 
simulations are compared to experimental jet data measured by the CMS experiment 
in PbPb and pp collisions at a center-of-mass energy of 2.76 TeV. Though both scenarios 
qualitatively describe the shape and main physical features of the FFs, the ratios are 
in much better agreement with the first scenario. Results are presented for the Monte Carlo FFs 
obtained for different parton flavours (quark and gluon) and accounting exactly, or not, 
for the experimental jet reconstruction biases.

\end{titlepage}

\section{Introduction}

Experiments at the Relativistic Heavy Ion Collider (RHIC) and Large Hadron Collider 
(LHC) have observed the formation of a Quark-Gluon Plasma (QGP) in AuAu and PbPb collisions 
respectively. Highly virtual quarks and gluons 
(generically called partons) lose energy as they traverse the QGP, resulting 
in the suppression of high transverse momentum leading hadrons 
\cite{Adcox:2001jp,Adams:2003kv,Aamodt:2010jd,CMS:2012aa} and 
jets~\cite{Aad:2010bu,Chatrchyan:2011sx} as well as in the modification of 
jet fragmentation functions and jet shapes~\cite{Adare:2012qi,Chatrchyan:2013kwa,Chatrchyan:2014ava}, 
observed in central heavy-ion collisions.  

In the vacuum, the production of highly virtual partons issuing from a hard scattering 
of two partons from the incoming protons results in a spray of collimated hadrons 
observed in the final-state of the collision. The evolution of successive
splittings $q(\bar q)\to q(\bar q)g$, $g\to gg$ and $g\to q\bar q$ ($q$, $\bar q$ and $g$ 
label quark, antiquark and gluon respectively) inside the parton shower prior 
to hadronization can be computed analytically from perturbative QCD calculations resumming 
collinear and infrared divergences~\cite{Dokshitzer:1991wu}
or, alternatively, in terms of Monte Carlo (MC) formulations of the parton branching process such as 
the PYSHOW algorithm implemented in {\sc pythia} \cite{Bengtsson:1986hr,Norrbin:2000uu}. 
In nucleus-nucleus (A-A) collisions, partons produced in the hard scatterings of two partons 
from the nuclei propagate through the hot/dense QCD medium also produced in such collisions
and their branching pattern is changed by interacting with the color charges of the deconfined 
QGP \cite{Baier:2000mf}. As a consequence, additional medium-induced soft gluon 
radiation is produced in A-A collisions, which leads for instance to the suppression of
high-$p_T$ hadroproduction \cite{Salgado:2002cd,Gyulassy:2001nm,Wang:2002ri}
and a plethora of other jet modifications (see e.g.\cite{d'Enterria:2009am}). 
In the past few years, MC codes for in-medium shower simulations developed for hadronic
collisions have also become available~\cite{Zapp:2008gi,Renk:2008pp,Renk:2009nz,Armesto:2008zza,Armesto:2009fj}. 
They have been based on the success of MC shower simulations in the vacuum such as 
{\sc pythia} and {\sc herwig}~\cite{Corcella:2000bw}. 

The parton-to-hadron jet fragmentation functions (FFs), $\frac{dN}{d\xi}\equiv zD_{i\to h}(z,Q)$ with
$\xi=\ln(1/z)$, encode the probability that a parton $i$ fragments into a hadron $h$ carrying a fraction
$z$ of the parent parton's momentum. In this paper, we compute the medium-modified 
FFs and the FF ratio of the fragment yield with {\sc YaJEM}, 
where the medium itself is described by a 3-d hydrodynamical evolution~\cite{Renk:2008pp,Renk:2009nz}. 
We compare our results with recent PbPb and pp CMS data collected at center-of-mass energy 
2.76 TeV~\cite{Chatrchyan:2014ava}. In the first scenario of the {\sc YaJEM} code, 
it is mainly assumed that the cascade of branching 
partons traverses a medium which, consistently with standard radiative loss pictures,
is characterized by a local transport coefficient $\hat{q}$ which measures the virtuality
per unit length transferred from the medium to the leading parton. Hence,
the virtuality of the leading parton is increased by the integrated amount 
``$\Delta Q^2$" which opens up the phase space and leads to a softer shower. 
The second scenario is based on the Borghini-Wiedemann (BW) model~\cite{Borghini:2005em}, where
the singular part of the branching kernels in the medium is enhanced by a factor 
$1+f_{\rm med}$, such that $P_{a\to bc}=(1+f_{\rm med})/z+{\cal O}(1)$,
where $a\to bc$ describes the possible QCD parton branchings, i.e. $q(\bar q)\to q(\bar q)g$ and 
$g\to gg$ with $g\to q\bar q$ unchanged. In this case, the softening of the shower is described
by the larger amount of medium-induced soft gluons ($f_{\rm med}>0$) as compared to the vacuum ($f_{\rm med}=0$). 
In both scenarios, the final parton-to-hadron transition takes place in the vacuum, 
using the Lund model~\cite{Andersson:1983ia}, for hadronization scales below $Q_0=1$ GeV. 

For the purpose of a realistic comparison of {\sc YaJEM} and {\sc pythia6} with the CMS data, 
the FF analysis is carried out by following the 
CMS analysis closely. Jets are reconstructed with the anti-$k_t$ 
algorithm \cite{Cacciari:2011ma,Cacciari:2005hq} with a 
resolution parameter $R=0.3$. The clustering analysis is 
limited to charged particles with $p_{t}>1$ GeV inside the jet cone where 
$P_T\geq100, 120, 150$ GeV are required for jets (i.e. $P_T$ stands for the 
jet transverse momentum inside the jet cone after reconstruction) in the jet $P_T$ ranges 
$100\leq P_T({\rm GeV})\leq 120$, $120\leq P_T({\rm GeV})\leq 150$, 
$150\leq P_T({\rm GeV})\leq 300$ and $100\leq P_T({\rm GeV})\leq 300$ reported 
by the CMS collaboration~\cite{Chatrchyan:2014ava}. The 
condition $p_{t}>1$ GeV facilitates the experimental jet reconstruction 
(as it removes a very large underlying-event background) but can 
potentially bias the jet FF analysis. In order to illustrate the role of the bias caused by the jet-finding procedure, 
we compare the biased FFs (i.e. provided the CMS jet-finding conditions are fulfilled) with
the unbiased FFs (i.e. the ones obtained by analyzing all jets, including 
the ones found close to their nominal energy) for both PbPb and pp CMS data.

\section{Comparison with {\sc YaJEM} and QGP hydrodynamics}

\label{sec:yajemdescription}

In the vacuum, the PYSHOW algorithm \cite{Bengtsson:1986hr,Norrbin:2000uu} is a 
well-tested numerical implementation of QCD shower simulations. {\sc YaJEM} is based on the PYSHOW algorithm, 
to which it reduces in the limit of no medium effects. It simulates the evolution of 
a QCD shower as an iterated series of splittings of a parent into two offspring partons 
$a \rightarrow  bc$ where the energy of the offspring are obtained as $E_b = z E_a$ 
and $E_c = (1-z) E_a$ and the virtuality of parents and offspring is ordered 
as $Q_a \gg Q_b, Q_c$. In the explicit kinematics of the MC shower, the singularities at 
$z\rightarrow 0$ or $z \rightarrow 1$ lie outside of the accessible phase space and no $[\dots]_+$ 
regularization procedure is needed. The  decreasing hard virtuality scale of partons provides, 
splitting by splitting, the transverse phase space for radiation, and the perturbative 
QCD evolution terminates once the parton virtuality reaches the value $Q_0 = 1$ GeV,  
followed by the hadronization using the Lund model \cite{Andersson:1983ia}.

The {\sc YaJEM} scenario (i) makes the assumption that the  virtuality of partons traversing 
the medium grows according to the medium transport coefficient $\hat{q}(\zeta)$ which measures
the virtuality transfer per unit pathlength. This coefficient is taken proportional to 
\begin{equation}\label{eq:qhat}
\hat{q}(\zeta)=K\cdot2\cdot\epsilon^{3/4}(\zeta)F(\rho(\zeta),\alpha(\zeta))
\end{equation}
with
$$
F(\rho(\zeta),\alpha(\zeta))=\cosh\rho(\zeta)-\sinh\rho(\zeta)\cos\alpha(\zeta),
$$
where $\epsilon$ is the local energy density of the hydrodynamical medium, $F$ is a hydrodynamical 
flow correction factor accounting for the Lorentz contraction 
of the scattering centers density as seen by the hard parton for $\rho(\zeta)$, which is 
the local flow rapidity and $\alpha(\zeta)$, the  angle between the hydrodynamical
flow and the parton propagation direction. 

For a shower parton $a$, created at a time $\tau^0_a$ and evolving during $\tau_a$ before
branching into a pair of offspring partons, the integrated virtuality as propagated inside 
the shower code is given by
\begin{equation}\label{eq:deltaq2}
\Delta Q^2=\int_{\tau^0_a}^{\tau^0_a+\tau_a}d\zeta\hat{q}(\zeta), 
\end{equation}
which increases the phase space from $Q^2\to Q^2+\Delta Q^2$ and thereby, the probability for 
medium-induced radiation. The integration in Eq.~(\ref{eq:deltaq2}) is performed over the eikonal 
trajectory of the parton-initiated shower from the production vertex to the boundary of the medium (see below).
Note that this scenario is different from {\sc YaJEM-E} and the  
{\sc YaJEM-DE}~\cite{Renk:2014lza}~\cite{Renk:2009nz} where the shower should be 
evolved down to $Q_0=\sqrt{E/L}$ (i.e. $E$ is the parton's energy and $L$, the medium length).

In the second scenario (BW model), the QCD splitting functions are enhanced in the infrared sector
according to the form~\cite{Borghini:2005em},
\begin{equation}
P_{q\to qg}=\frac43\frac{1+z^2}{1-z}\Rightarrow\frac43\left(\frac{2(1+f_{\rm med})}{1-z}-(1+z)\right)
\end{equation}
This increased branching probability leads to additional medium induced soft gluon production which
decreases the jet energy collimation and widens the jet-shape~\cite{PerezRamos:2012ci,Ramos:2014mba}, 
although no explicit flow of momentum between jet and medium is modeled. 
In particular, from the point of view of the leading parton, this
is a fractional energy loss mechanism since it is formulated as a function of the splitting variable $z$
only: the lost average energy due to the medium effect is proportional to the initial energy of the 
leading parton. The factor $f_{\rm med}$ is assumed to be proportional to
\begin{equation}
\label{E-fmed}
f_{med}=K_f\cdot\int d \zeta [\epsilon(\zeta)]^{3/4} F(\rho(\zeta),\alpha(\zeta)).
\end{equation}
We will refer to the implementation of the BW prescription for in-medium showers in the following as 
{\sc YaJEM+BW}. This is distinct from the version of the {\sc YaJEM} code described above, 
but also from {\sc YaJEM-E} and {\sc YaJEM-DE} which are tested against a large body of observables at both 
RHIC and the LHC (see e.g. Refs.~\cite{Renk:2011aa,Renk:2012cb,Renk:2012hz}).

In a MC treatment of the shower evolution, using a fixed value for $\Delta Q^2$ or $f_{\rm med}$ 
to characterize the medium is not needed and is in fact not realistic once a comparison with the data is desired. 
Following the procedure described in Ref.~\cite{Renk:2009nz}, the values of $\Delta Q^2$ and $f_{\rm med}$ for both scenarios 
are determined event-by-event by embedding the hard process into a hydrodynamical medium \cite{Renk:2011gj} 
starting from a binary vertex at $(x_0,y_0)$, and evaluating the line integral in Eq.~(\ref{eq:deltaq2}) 
and Eq.~(\ref{E-fmed}) over the eikonal trajectory $\zeta$ through the medium. 
Events are then generated for a large number of random $(x_0,y_0)$ sampled from the 
transverse overlap profile
\begin{equation}
\label{E-Profile}
P(x_0,y_0) = \frac{T_{A}({\bf r_0 + b/2}) T_A(\bf r_0 - b/2)}{T_{AA}({\bf b})},
\end{equation}
where $T_{A}$ is a nuclear thickness function $T_{A}({\bf r})=\int dz \rho_{A}({\bf r},z)$ 
obtained from the Woods-Saxon density $\rho_{A}({\bf r},z)$ and $b$ is the impact parameter. 
All observables are averaged over a sufficiently large number of events. This procedure leaves two 
dimensionful parameters $K$ in Eq.~(\ref{eq:qhat}) and $K_f$ in Eq.~(\ref{E-fmed}) 
characterizing the strength of the coupling between parton and 
medium which are tuned to reproduce the measured nuclear suppression 
factor $R_{AA}$ for each scenario 
in central 200 GeV AuAu collisions (see  Ref.~\cite{Renk:2009nz}).

\section{Monte Carlo analysis of medium-modified FFs}
 
{\sc YaJEM} and {\sc pythia6} generate back-to-back showers at center-of-mass energy $\sqrt{s}$.
In order to simulate the huge amount of jets produced in pp (vacuum) and PbPb 
(medium) in central collisions ($10\%$), we need to compute the initial $p_T$ 
distribution of partons before the showering process and its interaction with the 
medium start (i.e. $p_T$ stands for the parton transverse momentum and differs from 
the jet $P_T$ before clustering and accounting for cuts). In the vacuum, 
the initial distribution of gluon and quark-initiated showers
is determined by the convolution of parton distribution functions (PDFs) and the 
leading order (LO) matrix elements of the hard scattering cross section at the 
given factorization scale of the hard process. For the medium, the same calculation
is repeated with the nuclear parton distribution functions (nPDFs).
PDFs and nPDFs are provided by the CTEQ \cite{Lai:2010nw} 
and EKS \cite{Eskola:1998df} families for pp and PbPb collisions 
in the vacuum and the medium respectively. In both cases, the initial 
distribution can be approximated by a fast decreasing power low 
like $(1/p_T)^{\alpha}$, which is different for RHIC and the LHC.

Since partons are copiously produced in the LHC environment at center-of-mass-energy 2.76 TeV, 
we sample the initial distribution of partons described above 
by randomly selecting two thousand gluon and quark dijets with center-of-mass energy
$\sqrt{s}\sim 2p_T$ as input to {\sc pythia6} (vacuum) and {\sc YaJEM} (medium) 
over each $p_T$ range. Jets are clustered by using the anti-$k_t$ algorithm for each $p_T$ range inside 
the jet cone of resolution $R=0.3$ with charged particles as in the CMS 
experiment. Note that we purposely use the default algorithm currently used by all LHC experiments.
Our motivation to do so relies on the fact that the anti-$k_t$ is the most robust jet 
reconstruction algorithm for pp and PbPb collisions at the LHC with respect to 
underlying events and pileup.

Reconstructed jets can be sorted by $P_T$ ($P_{T1}>P_{T2}>\ldots$) for the analysis
such that the most hardest one ($P_{T1}$) can be randomly selected from its pair ($P_{T2}$) 
event-by-event. Since the correlation between initial parton kinematics ($p_T$) 
and reconstructed jet kinematics ($P_T$) gets increasingly blurred for small 
reconstruction radii ($R=0.3$) and soft background removal ($p_{t}>1$ GeV), 
the transverse momentum $P_T$ of various jets 
on each sample can drop below the original $p_T$ range of the leading parton (i.e. $p_T<P_T$). 
Since the CMS data for FFs and the PbPb/pp ratio of FFs spans over four jet $P_T$ ranges:  
$100\leq P_T({\rm GeV})\leq 120$, $120\leq P_T({\rm GeV})\leq 150$, 
$150\leq P_T({\rm GeV})\leq 300$ and $100\leq P_T({\rm GeV})\leq 300$~\cite{Chatrchyan:2014ava}, 
a second $P_T$ filtering is required in order to fulfill CMS trigger bias conditions, 
such that $P_T\geq100$ GeV, $P_T\geq120$ GeV, $P_T\geq150$ GeV and $P_T\geq100$ GeV respectively. 
The requirement imposed by the trigger selection in the analysis is defined as ``biased", 
while that including jets with all clustered hadrons is called ``unbiased". 

For this reason, the fraction of gluon 
jets in one sample is decreased by the trigger bias from its theoretical (unbiased) 
value $f_g^{\rm u, vac}$ to $f_g^{\rm b, vac}$ for biased showers in the vacuum, and from 
$f_g^{\rm u, med}$ to $f_g^{\rm b, med}$ for biased showers in the medium. Their values 
depend weakly on the chosen medium-induced radiation scenario. The 
values of $f_g^{\rm u, vac}$ and $f_g^{\rm u, med}$ can be determined from the initial
distribution of gluons in each $P_T$ range, while $f_g^{\rm b, vac}$ and 
$f_g^{\rm b, med}$ can be obtained after computing the fraction of gluon jets  
passing the $P_T$ trigger selection. For the sake of a realistic comparison with the CMS 
data, one should evaluate the mixed FFs which are obtained from the linear combinations 
of gluon and quark FFs:
\begin{equation}\label{eq:mixedFF}
\left(\frac{dN}{d\xi}\right)_{\rm mixed}=f_g\left(\frac{dN}{d\xi}\right)_{g}
+(1-f_g)\left(\frac{dN}{d\xi}\right)_{q}.
\end{equation}
For the simulation procedure, we perform a double random selection of events over $p_T$ and $\Delta Q^2$ 
($f_{\rm med}$) for the first (second) scenario as input to {\sc YaJEM} ({\sc YaJEM+BW}) and average  
over both variables for a large number of events. In the first scenario, the {\sc YaJEM} code is run 
by averaging over $p_T$ and the medium-induced virtuality $\Delta Q^2$, which results in 
$(\langle P_T^{\rm u}\rangle$, $\langle\Delta Q^2\rangle)$ and in the second one, it is run 
as {\sc YaJEM+BW} by averaging over $p_T$ and $f_{\rm med}$, which results in 
$(\langle P_T^{\rm u}\rangle$, $\langle f_{\rm med}\rangle)$. 
Such average values are obtained after embedding the hard process in the hydrodynamical medium, 
as described in section~\ref{sec:yajemdescription}. In such a way, the trigger 
selection can be applied to the jet $P_T$ and medium parameter simultaneously after 
clustering, resulting in $(\langle P_T^{\rm b}\rangle$, $\langle\Delta Q^2\rangle)$ and 
$(\langle P_T^{\rm b}\rangle$, $\langle f_{\rm med}\rangle)$ 
(i.e. $\langle P_T^{\rm u}\rangle$ stands for the average $P_T$ value over all events 
after clustering and $\langle P_T^{\rm b}\rangle$ is the  
average $P_T$ of those jets fulfilling the trigger bias selection.)

\subsection{Medium-modified FFs and ratio: Comparison with CMS data at 2.76 TeV}

In this subsection we proceed to compare the CMS PbPb, pp FFs and the ratio 
PbPb/pp of FFs with {\sc YaJEM}, {\sc pythia6} and {\sc YaJEM/pythia6} versus
{\sc YaJEM+BW/pythia6} respectively. We note that since the FFs written as a function of $\xi\equiv\ln(1/z)$ 
(i.e $z=p_t/p_T$) follow a hump-backed plateau shape, such a shape can be parametrized, without any loss of generality,
as a distorted Gaussian (DG) which depends on the original $p_T$ of the parton in the range
$0\leq\xi\leq Y$ with $Y = \ln({p_T/Q_{_{0}}})$, evolved down to a shower cut-off scale 
$\lambda = \ln(\rm Q_{_{0}}/\lqcd)$:
\begin{equation}
\hspace{-0.7cm}D(\xi,Y,\lambda) = {\cal N}/(\sigma\sqrt{2\pi})\cdot e^{\left[\frac18k-\frac12s\delta-
\frac14(2+k)\delta^2+\frac16s\delta^3+\frac1{24}k\delta^4\right]}\,, 
\label{eq:DG}
\end{equation}
where $\delta=(\xi-\bar\xi)/\sigma$, with moments:
${\cal N}$ (hadron multiplicity inside the jet), $\bar\xi$ (DG peak position), 
$\sigma$ (DG width), $s$ (DG skewness), and $k$ (DG kurtosis). The energy-evolution of these moments 
can be analytically calculated at NNLL+NLO* accuracy and compared to the jet data to extract a 
very precise value of the strong QCD coupling $\alpha_s$~\cite{d'Enterria:2014bsa,d'Enterria:2014yya,Perez-Ramos:2013eba}
in elementary $e^+e^-$ and $e^-p$ collisions. In a forthcoming work~\cite{denterria2015}, 
we plan to generalize such an approach to heavy-ion phenomenology in order to the extract the 
medium transport coefficient $\hat{q}$ and pathlength $L$.  

In Table~\ref{table:reccoeff} we display the gluon fractions of the biased and unbiased 
showers for each $p_T$ range, as well as the unbiased $\langle P_T^{\rm u}\rangle$ and biased 
$\langle P_T^{\rm b}\rangle$ transverse momenta.

\begin{table}[htb]
\begin{center}
\begin{tabular}{cccccccc}
\hline\hline
$p_T$ range (GeV) & $\langle P_T^{\rm u}\rangle$ (GeV)
& $\langle P_T^{\rm b}\rangle$ (GeV) & $f_g^{\rm u, vac}$ & $f_g^{\rm b, vac}$ & $f_g^{\rm u, med}$ & $f_g^{\rm b, med}$ \\ \hline
100-120 & 44.6 & 103.0 & 0.463 & $5.2\times10^{-4}$ & 0.330 & $1.1\times10^{-4}$ &\\
120-150 & 55.9 & 124.5 & 0.399 & $1.1\times10^{-3}$ & 0.182 & $1.3\times10^{-4}$ &\\
150-300 & 97.2 & 173.8 & 0.376 & $5.1\times10^{-2}$ & 0.174 & $1.5\times10^{-2}$ &\\
100-300 & 85.8 & 134.6 & 0.463 & 0.168 & 0.330 & 0.096 &\\
\hline\hline
\end{tabular}
\caption{Recovered jet transverse momentum and gluon fractions inside the jet cone $R=0.3$.}
\label{table:reccoeff}
\end{center}
\end{table}

\begin{figure}[!htbp]
\begin{center}
\epsfig{file=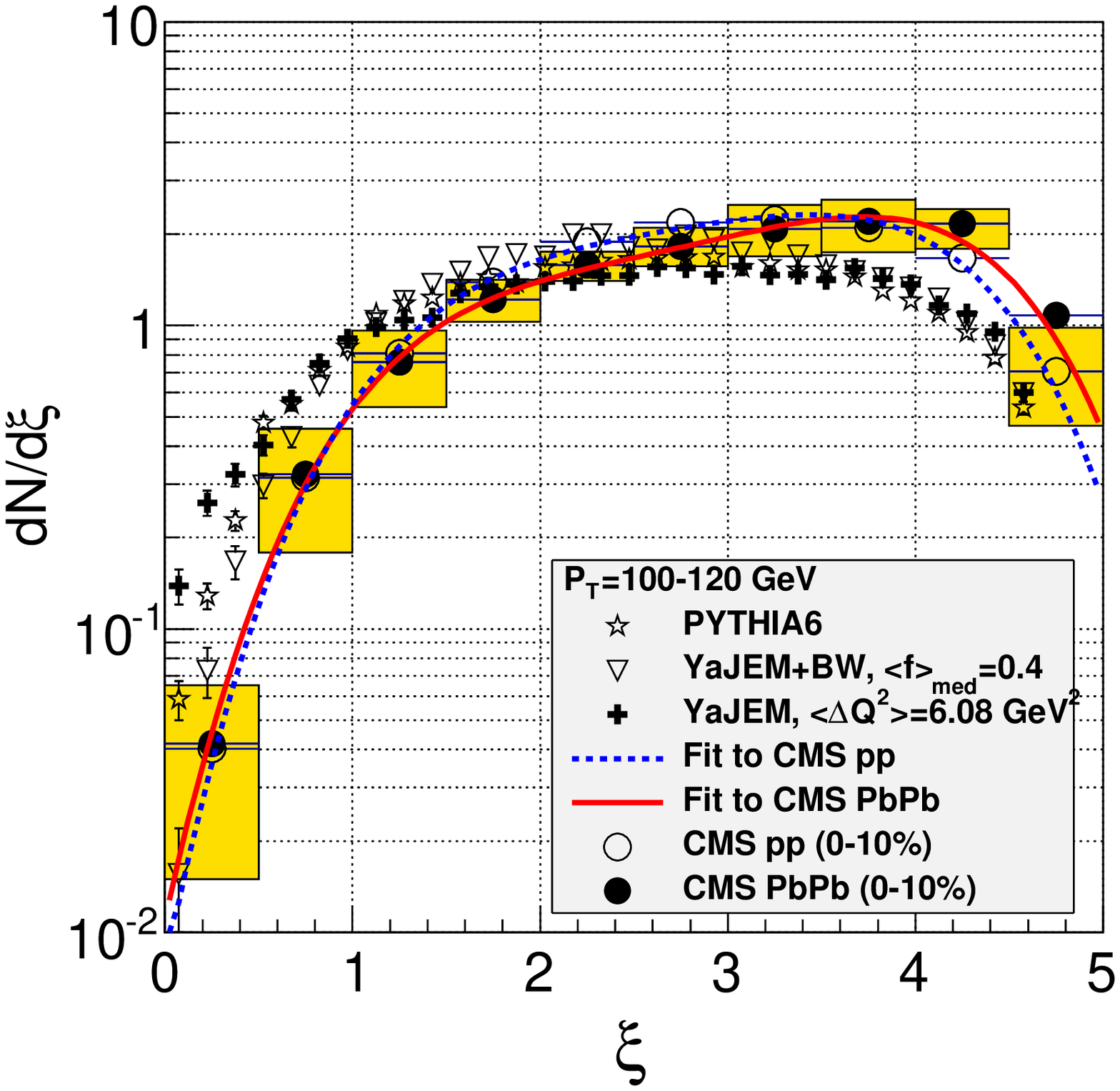, height=8.0truecm,width=8.4truecm}
\epsfig{file=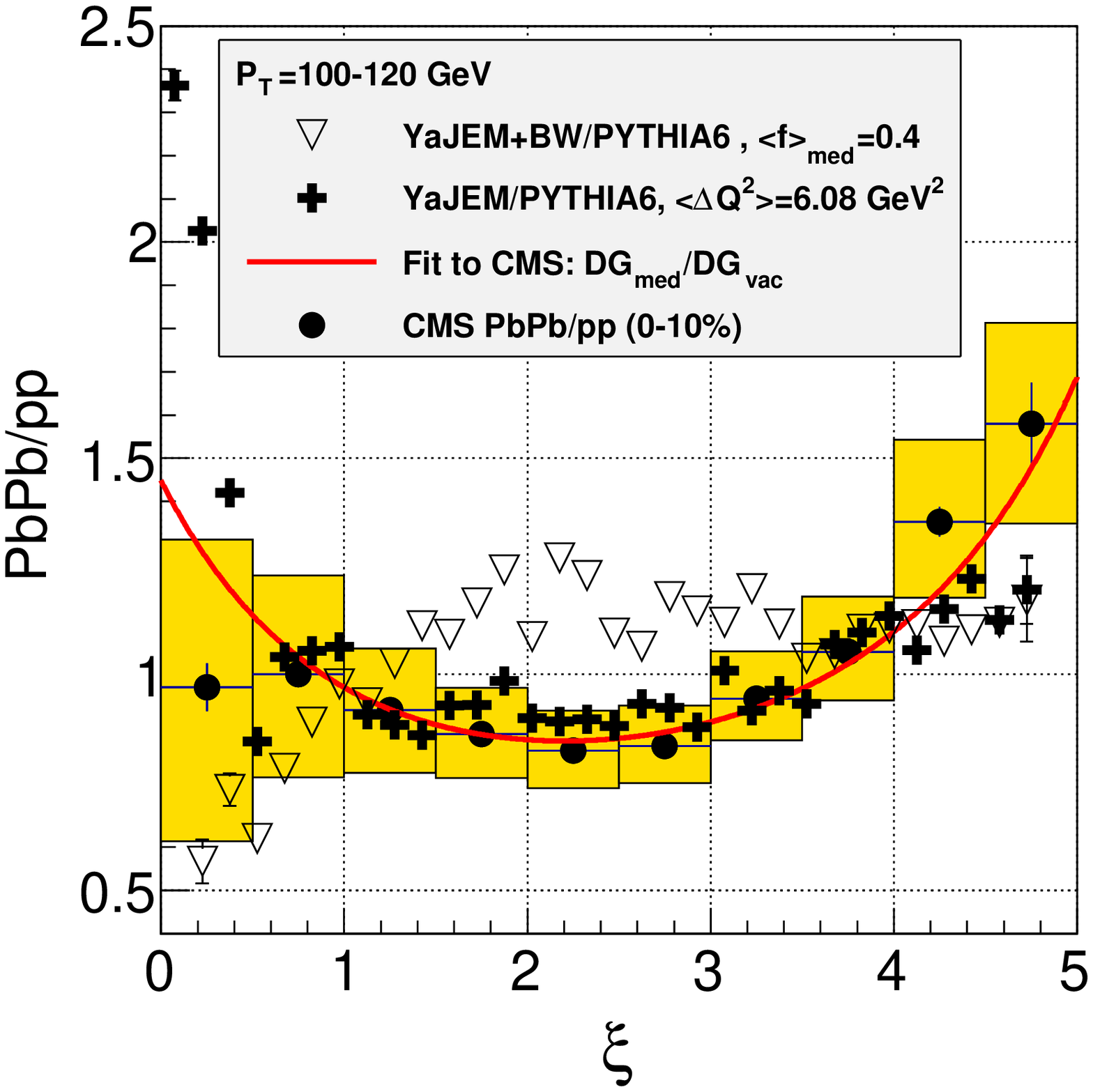, height=8.0truecm,width=8.4truecm}
\caption{\label{fig:ff100_120} Comparison of jet fragmentation 
functions in pp and PbPb collisions, for jets with $100\leq P_T ({\rm GeV})\leq120$, 
measured by CMS~\cite{Chatrchyan:2014ava} and obtained in two MC 
approaches ({\sc YaJEM} and {\sc YaJEM+BW}): absolute 
distributions (left), and PbPb/pp ratios (right).}  
\epsfig{file=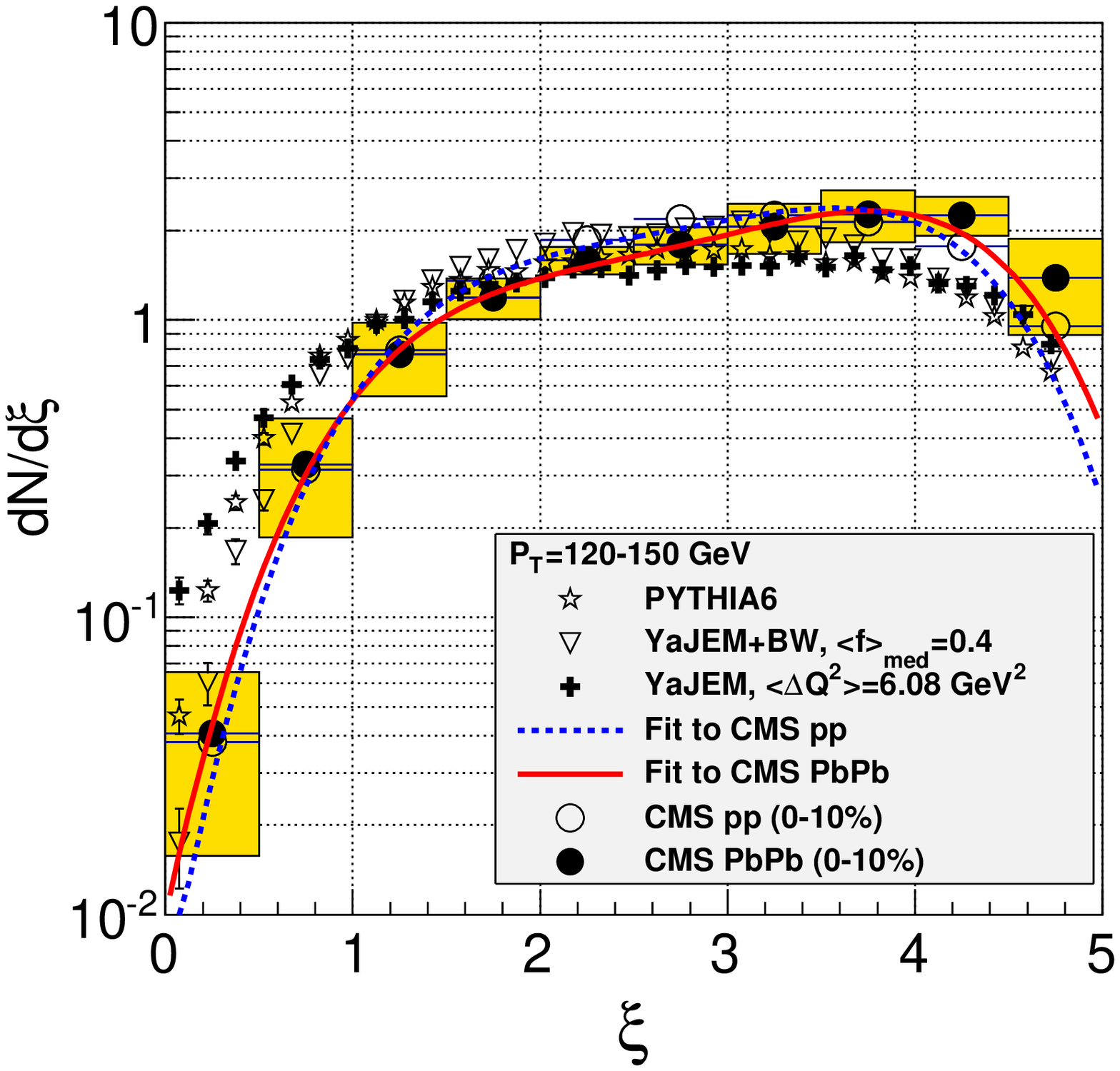, height=8.0truecm,width=8.4truecm}
\epsfig{file=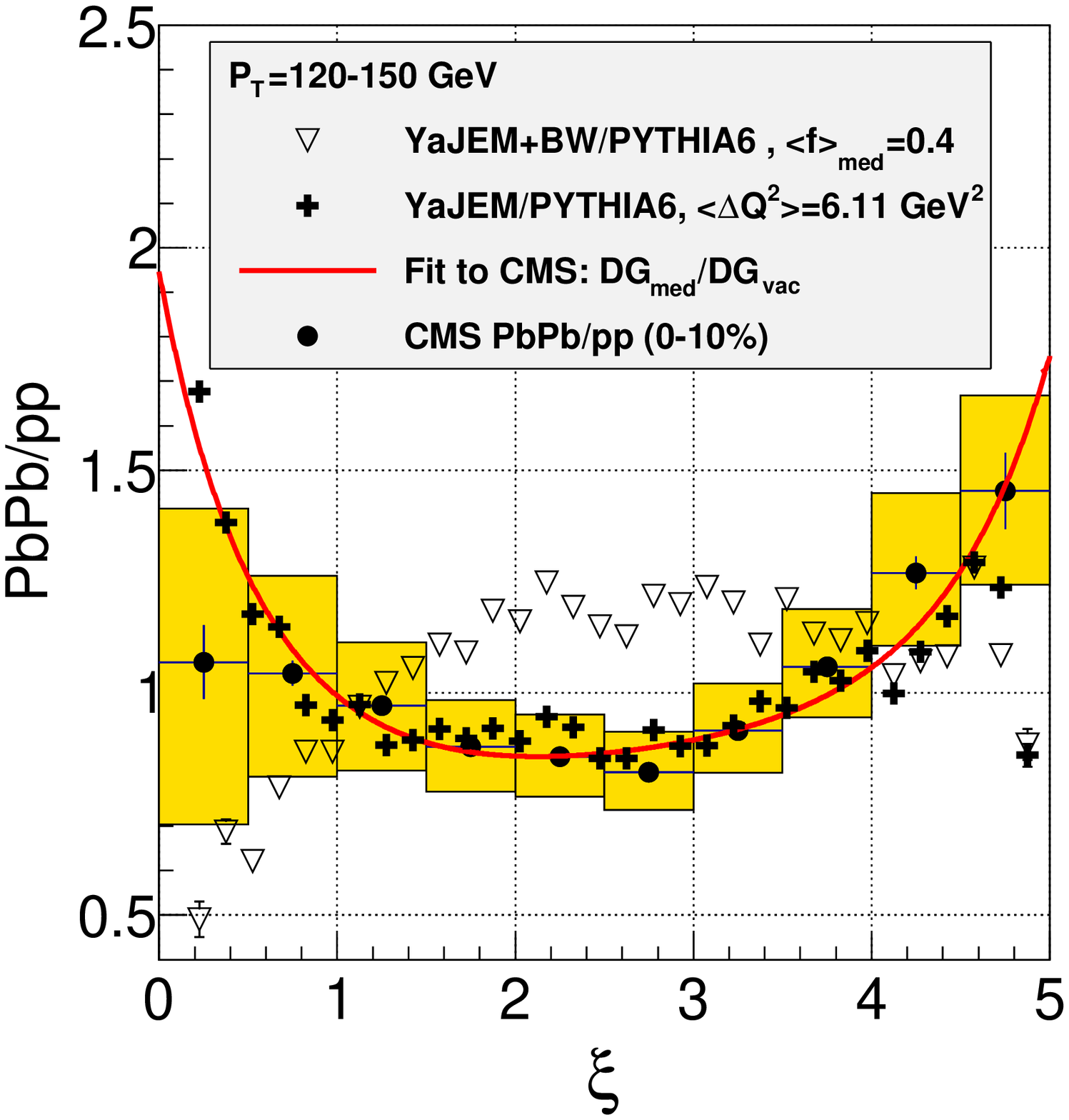, height=8.0truecm,width=8.4truecm}
\caption{\label{fig:ff120_150} Comparison of jet fragmentation 
functions in pp and PbPb collisions, for jets with $120\leq P_T ({\rm GeV})\leq150$, 
measured by CMS~\cite{Chatrchyan:2014ava} and obtained in two MC 
approaches ({\sc YaJEM} and {\sc YaJEM+BW}): absolute 
distributions (left), and PbPb/pp ratios (right).}  
\end{center}
\end{figure}

\begin{figure}[!htbp]
\begin{center}
\epsfig{file=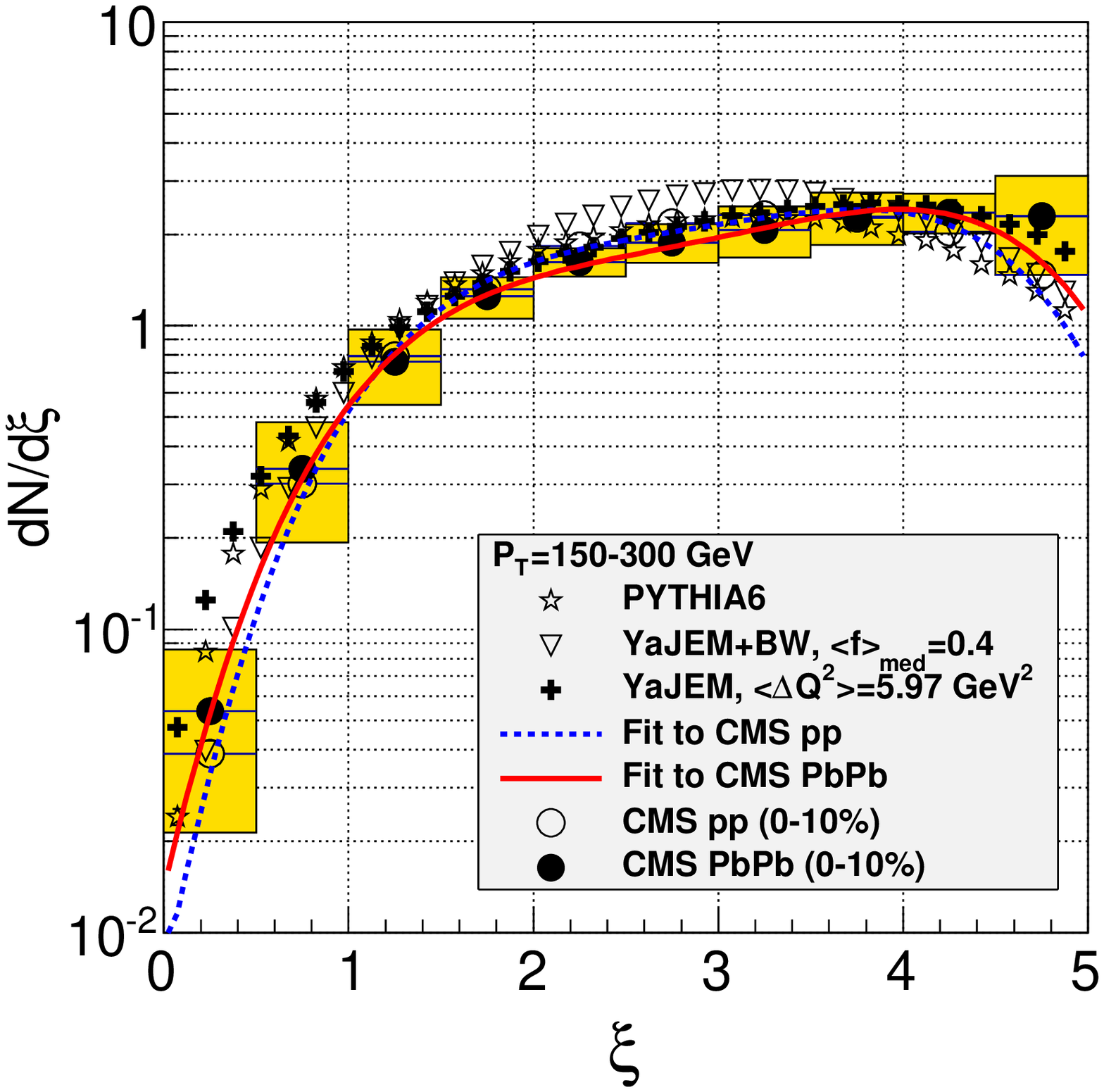, height=8.0truecm,width=8.4truecm}
\epsfig{file=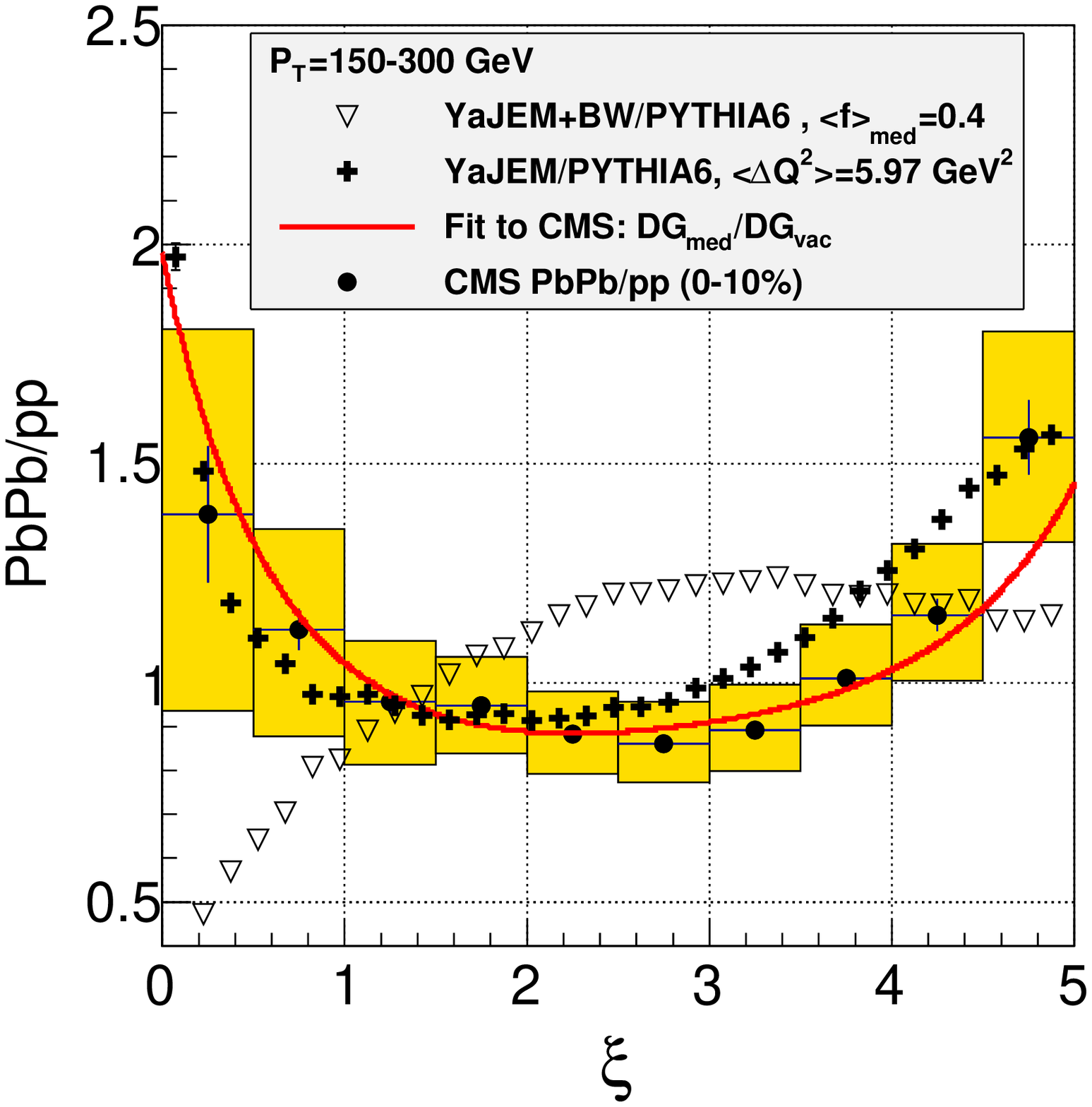, height=8.0truecm,width=8.4truecm}
\caption{\label{fig:ff150_300} Comparison of jet fragmentation 
functions in pp and PbPb collisions, for jets with $150\leq P_T ({\rm GeV})\leq300$, 
measured by CMS~\cite{Chatrchyan:2014ava} and obtained in two MC 
approaches ({\sc YaJEM} and {\sc YaJEM+BW}): absolute 
distributions (left), and PbPb/pp ratios (right).}  
\epsfig{file=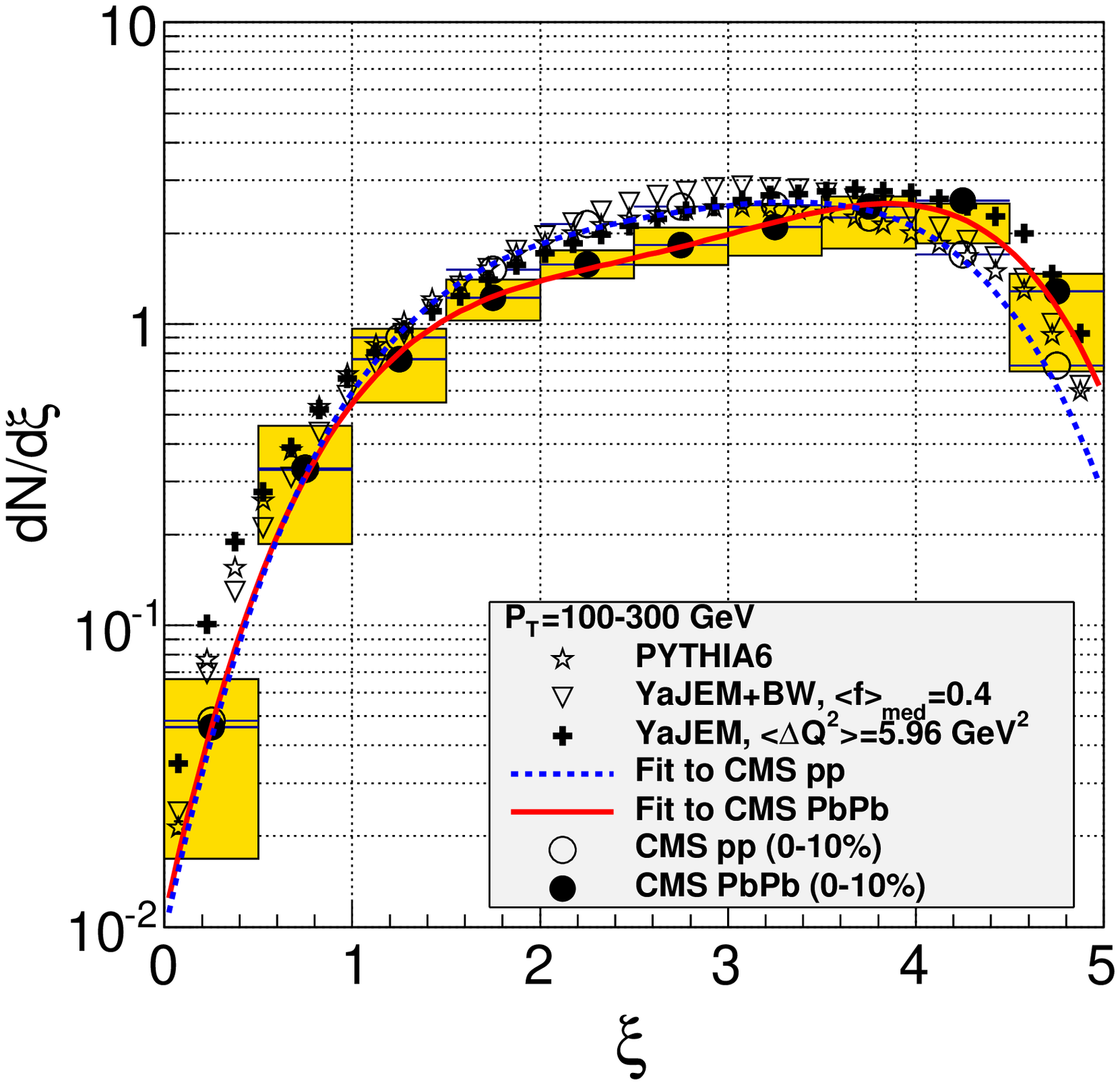, height=8.0truecm,width=8.4truecm}
\epsfig{file=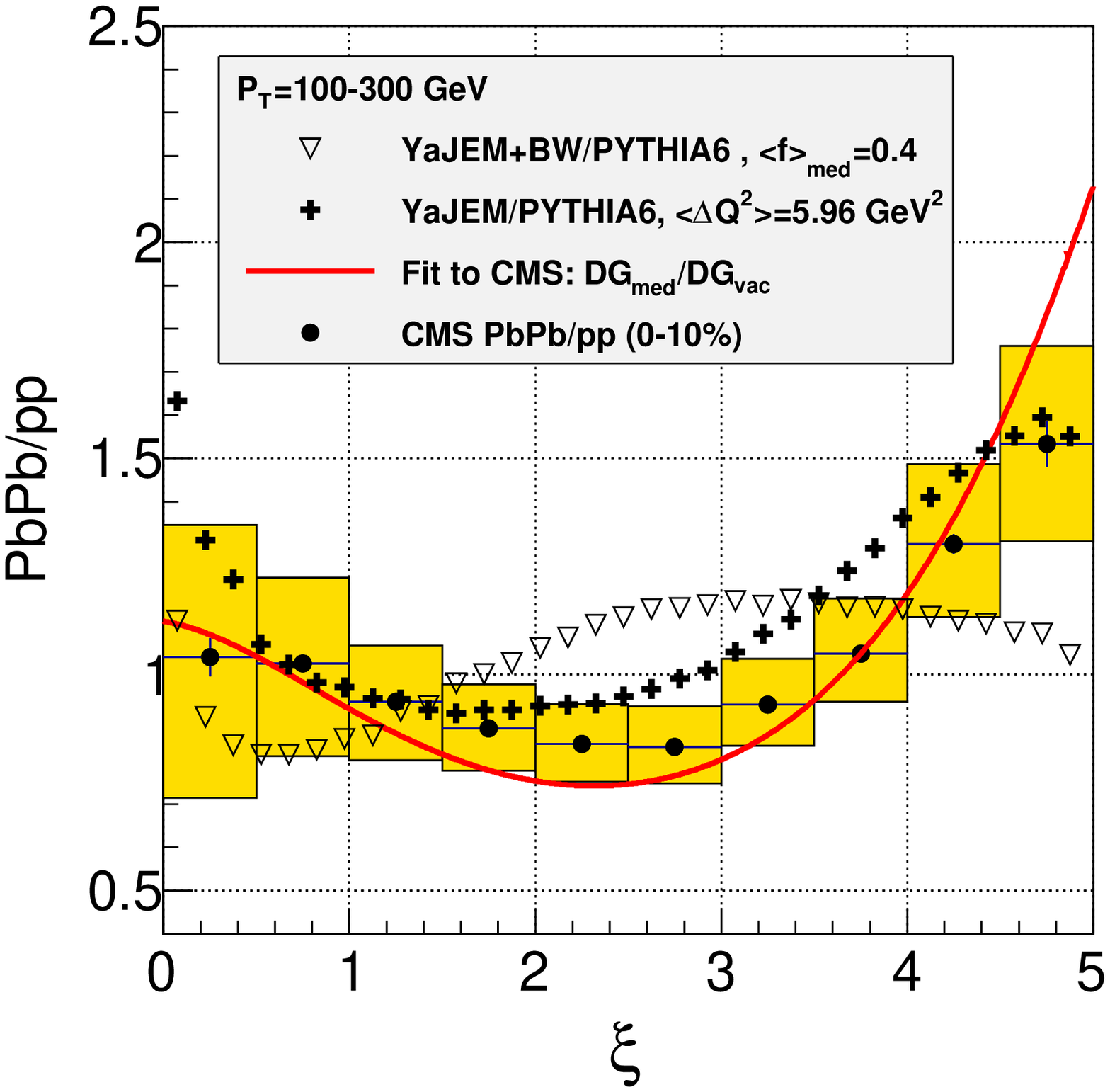, height=8.0truecm,width=8.4truecm}
\caption{\label{fig:ff100_300} Comparison of jet fragmentation 
functions in pp and PbPb collisions, for jets with $100\leq P_T ({\rm GeV})\leq300$, 
measured by CMS~\cite{Chatrchyan:2014ava} and obtained in two MC 
approaches ({\sc YaJEM} and {\sc YaJEM+BW}): absolute 
distributions (left), and PbPb/pp ratios (right).}  
\end{center}
\end{figure}

We display in Figs.~\ref{fig:ff100_120}--\ref{fig:ff100_300} the PbPb and pp FFs, 
and their ratios, for various jet $P_T$ ranges at 2.76 TeV comparing the CMS results 
to the PYTHIA6, YAJEM and YAJEM+BW predictions. In order to highlight the trends 
of the data in the plots, we fit the pp and PbPb data points to the DG given
by Eq.(\ref{eq:DG}) and produce the ratios ${\rm DG}_{\rm med}/{\rm DG}_{\rm vac}$. 
The FF $\xi$ interval is determined by $\langle P_T^{\rm b}\rangle$ for each $p_T$ range and 
turns to be $0\leq\xi\lesssim5$. However, since {\sc YaJEM} is known to fail for 
very soft hadroproduction from 2 GeV, we limit the phenomenological discussion to the region
$0\leq\xi\lesssim4.3$, where results are expected to be robust.
We make use of Eq.(\ref{eq:mixedFF}) for mixed 
samples of gluon and quark jets and take the fractions of gluon 
jets displayed in Table~\ref{table:reccoeff} for the computation. The FFs for the data ranges 
$100\leq P_T({\rm GeV})\leq 120$ and $120\leq P_T({\rm GeV})\leq 150$ as displayed by Figs.~\ref{fig:ff100_120}
and \ref{fig:ff120_150} show an offset in the comparison with the CMS data in both scenarios, 
although they fall within the range of CMS uncertainties over all of the $\xi$ interval. 
However, we can conclude that the first scenario (i) provides a more accurate 
physical description of the medium-modified FF as confirmed 
by the ratios in the right panels of Figs.~\ref{fig:ff100_120} and \ref{fig:ff120_150}. 
Indeed, the first scenario ({\sc YaJEM}) reproduces the right concavity of the ratio for all $P_T$ 
ranges, while the trends in the second approach ({\sc YaJEM+BW}) show the opposite 
unphysical behavior and should be discarded from the phenomenological discussion hereafter. 
The best agreement for FFs in the comparison of the CMS data with the MC calculation is 
reached for the data ranges $150\leq P_T({\rm GeV})\leq 300$ and $100\leq P_T({\rm GeV})\leq 300$
in Figs.~\ref{fig:ff150_300} and \ref{fig:ff100_300} respectively. The first scenario is 
also here in much better agreement with the CMS PbPb data than that shown by the BW model. 
Likewise, {\sc pythia6} is in good agreement with pp data in both cases. Note that in all panels,
we have displayed the averaged values of the hydrodynamical-like parameters $\Delta Q^2$ and $f_{\rm med}$
which turn to be $\langle\Delta Q^2\rangle\sim6$ ${\rm GeV}^2$ and $\langle f_{\rm med}\rangle\sim0.4$. 
For a constant medium of length $L=2.5$ fm~\cite{Mehtar-Tani:2014yea}, the transport 
coefficient would be roughly $\hat{q}=\langle\Delta Q^2\rangle/L\sim2.4$ ${\rm GeV}^2/{\rm fm}$
according to this hydrodynamical prescription of the QGP.

Suppression of the hadron yield in the data is weak and can mainly be observed in 
the intermediate region around the maximum peak positions of FFs: 
$1.0\lesssim\xi\lesssim3.6$ ($0.4 \lesssim z\lesssim0.7$) for all data ranges 
in the right panels of Figs.~\ref{fig:ff100_120}--\ref{fig:ff100_300}. As a matter of fact, this is 
not surprising since, being the region where partons fragment more efficiently, they undergo 
more interactions with the medium and hence, a larger amount of them are dissipated inside the QGP 
before reaching the hadronization stage. For $0<\xi\lesssim1.0$ ($1<z\lesssim0.4$) 
and $3.8\lesssim\xi\lesssim4.3$ ($0.7 \lesssim z\lesssim 0.015$), hadroproduction is 
increased in PbPb compared to pp collisions, and particularly by a factor of 3/2 in 
the softest region $3.8\lesssim\xi\lesssim4.3$, as expected. 
However, for the ranges $100\leq P_T({\rm GeV})\leq 120$ and $120\leq P_T({\rm GeV})\leq 150$, 
this suppression property has been perfectly described by {\sc YaJEM}, 
while for $150\leq P_T({\rm GeV})\leq 300$ and $100\leq P_T({\rm GeV})\leq 300$, 
the region of soft hadroproduction is  slightly overestimated by {\sc YaJEM/pythia6} 
with a relative error of $\sim15\%$, but still within the CMS data uncertainties.  

\begin{figure}[!htbp]
\begin{center}
\epsfig{file=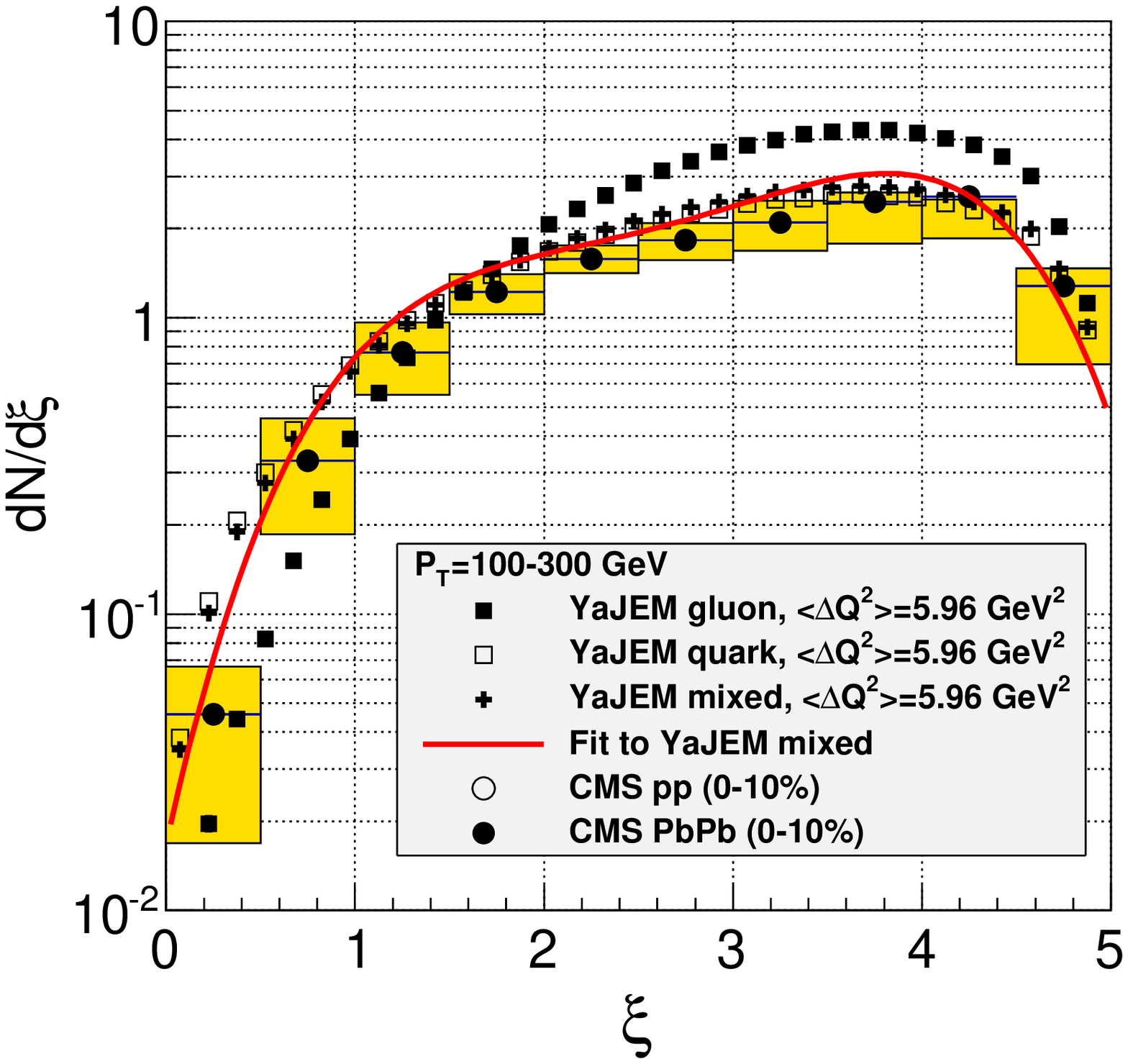, height=8.0truecm,width=8.4truecm}
\epsfig{file=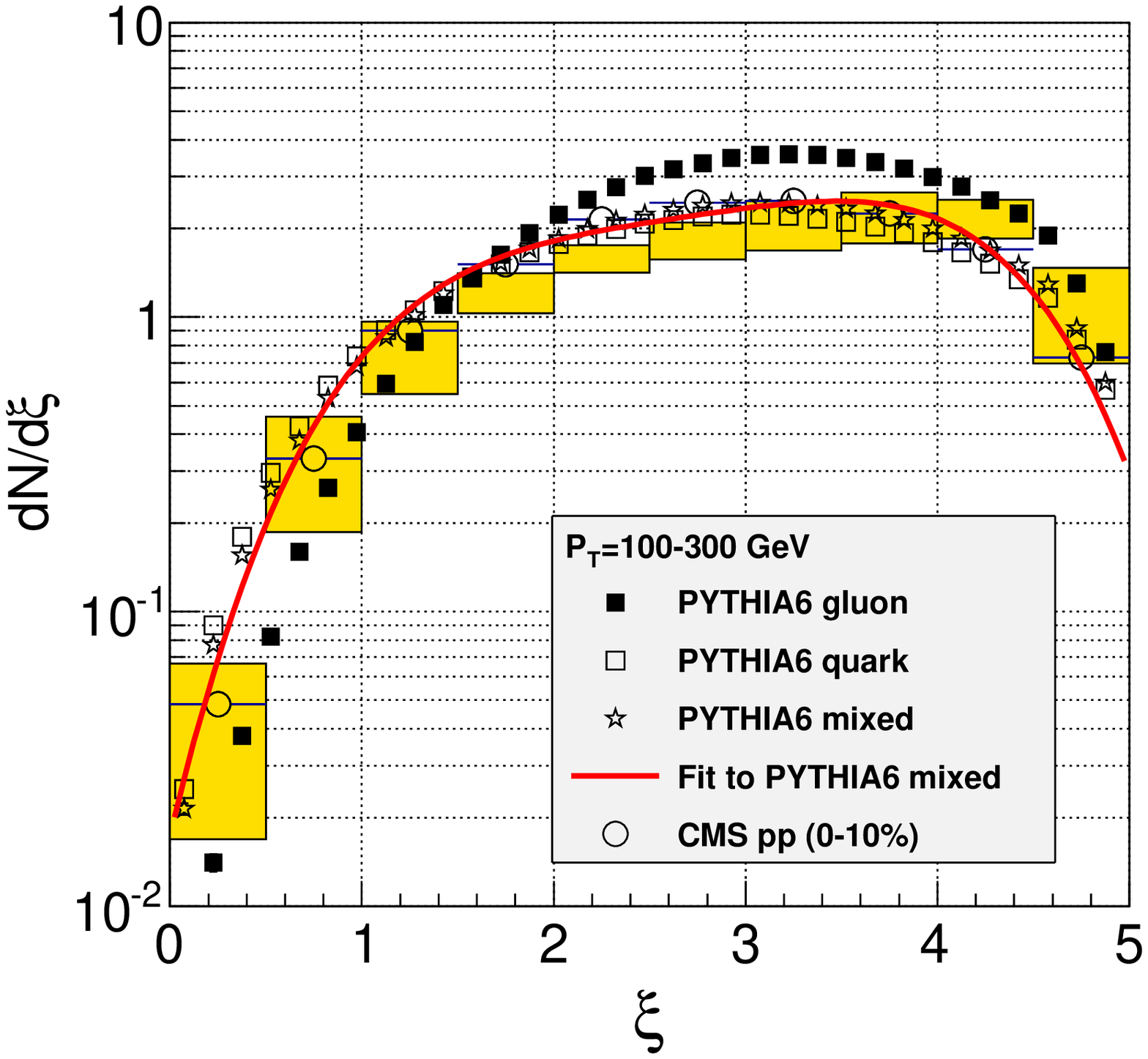, height=8.0truecm,width=8.4truecm}
\caption{\label{fig:quarkandgluon} Quark, gluon and parton fragmentation functions in PbPb (left) 
and pp (right) collisions obtained in the {\sc YaJEM} MC, 
for jets with $100\leq P_T ({\rm GeV})\leq300$, compared to CMS inclusive jet results~\cite{Chatrchyan:2014ava}.}  
\end{center}
\end{figure}

In Fig.~\ref{fig:quarkandgluon}, we compare the gluon and quark FFs obtained from {\sc pythia6}
and {\sc YaJEM} with the mixed FFs and CMS pp and PbPb data. Gluon jets produce a wider shower broadening 
than quark jets but they get even more suppressed by reconstruction biases than quark jets, which is clearly shown 
in Table~\ref{table:reccoeff}. In both cases, as expected, the quark FFs provided by {\sc pythia6}
and {\sc YaJEM} are in better agreement with the data. Similar trends for the 
ratio were found with an in-medium pathlength $L$-evolution as a consequence of gluon 
decoherence effects (anti-angular ordering) in the QGP~\cite{Mehtar-Tani:2014yea}.

\begin{figure}[!htbp]
\begin{center}
\epsfig{file=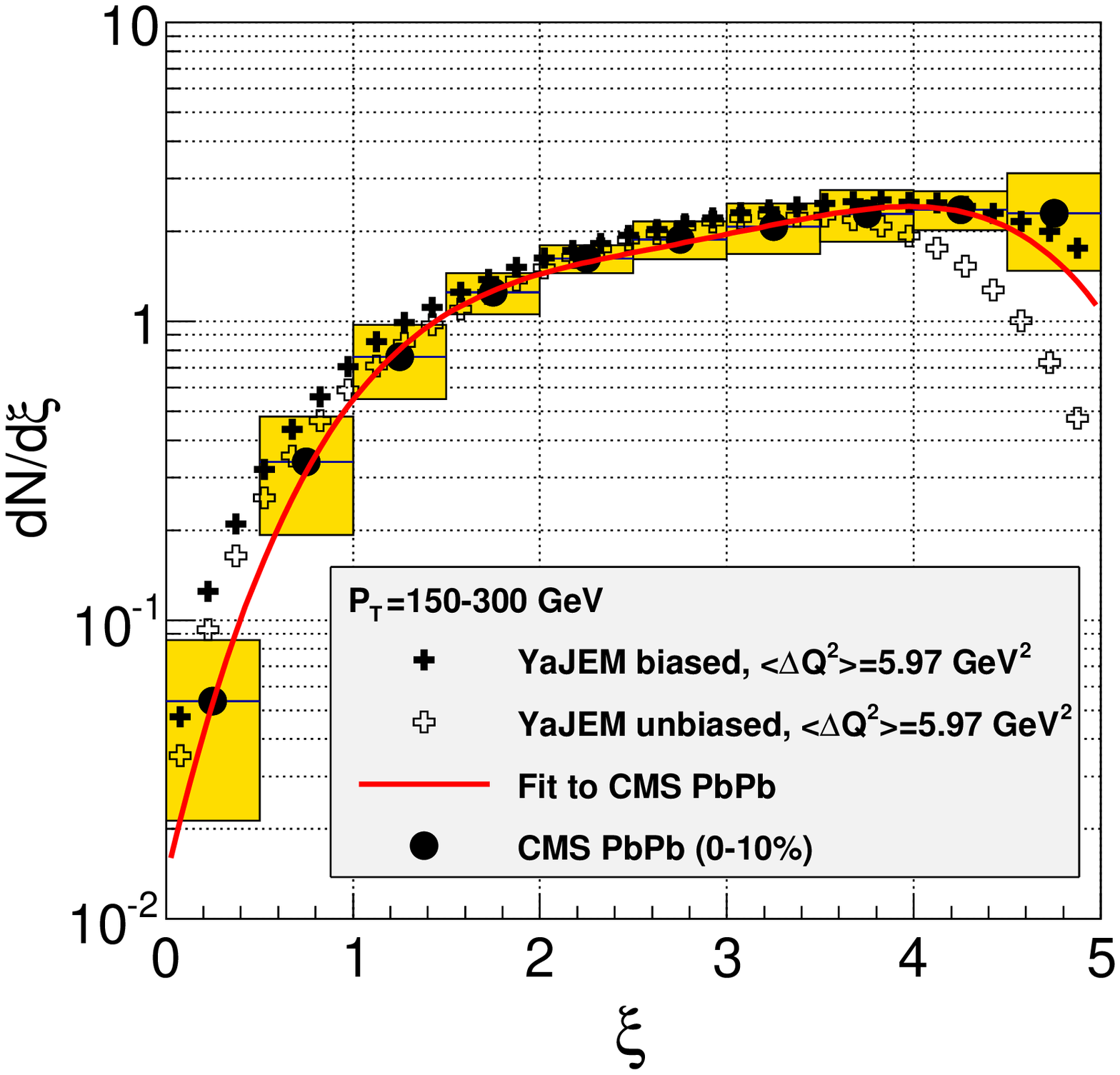, height=8.0truecm,width=8.4truecm}
\epsfig{file=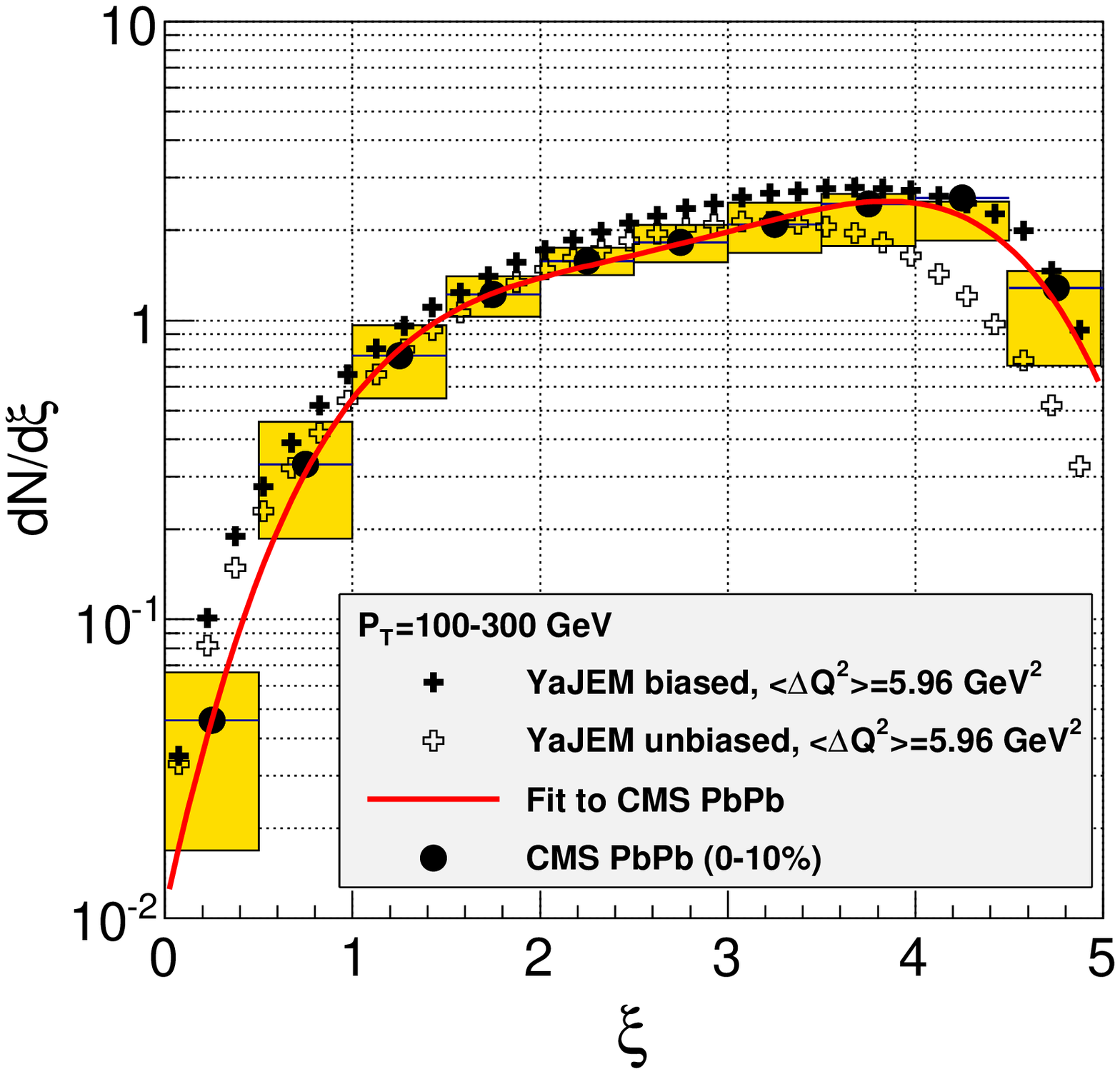, height=8.0truecm,width=8.4truecm}
\caption{\label{fig:biasedvsunbiased} Parton fragmentation functions in 
PbPb collisions in the {\sc YaJEM} Monte Carlo, 
for jets with $150\leq P_T ({\rm GeV})\leq300$ (left) and $100\leq P_T ({\rm GeV})\leq300$ 
(right) reconstructed applying the data-based cuts (``biased") or not (``unbiased"), 
compared to CMS inclusive jet results~\cite{Chatrchyan:2014ava}.}  
\end{center}
\end{figure}

For unbiased showers, the original fraction 
of gluon jets is relatively higher and should be taken from 
Table~\ref{table:reccoeff} for a direct comparison with biased showers in our MC 
study.
That is why we take the unbiased values for the fractions of gluon jets,
recompute Eq.~(\ref{eq:mixedFF}) in the given $P_T$ ranges and display the 
results in Fig.~\ref{fig:biasedvsunbiased} for $150\leq P_T({\rm GeV})\leq 300$ and 
$100\leq P_T({\rm GeV})\leq 300$. As illustrated, the unbiased shower shows an 
offset in the softest region $3.5\lesssim\xi\lesssim5$ compared to the data and the
biased FF simulated with {\sc YaJEM}. However, this bias is markedly higher 
for narrower $P_T$ ranges like $100\leq P_T({\rm GeV})\leq 120$ and 
$120\leq P_T({\rm GeV})\leq 150$ as displayed in Table~\ref{table:reccoeff}. 
 
The applicability of the Lund model in this framework relies on the fact 
that the hadronization of partons takes place in the vacuum. 
For a specific hadron $h$ of mass $m_h$, the spacial scale at which 
hadronization takes place is roughly $\sim E_h/m_h^2$. For kaons, protons and heavier 
hadrons this length can be drastically shortened. For pions however, which determine the
bulk of the multiplicity distribution in QCD showers, the Lund model can be safely applied 
such that the essential physical features and trends assumed by {\sc YaJEM} are still expected
to be consistent with the data down to $\sim2$ GeV. Its implementation in 
the PYSHOW algorithm mainly affects the tails of the FFs which in the 
softest region are widened compared to the parton shower without hadronization, but this effect is 
equally present in vacuum and medium showers as expected. 

\section{Summary}

In this paper, we studied the jet fragmentation functions (FFs) in pp and PbPb collisions, 
and their ratio, in two different Monte Carlo implementations of parton evolution in a 
quark-gluon plasma, compared with recent CMS PbPb and pp data at 2.76 TeV. 
The account of the medium-induced virtuality by the transport coefficient $\hat{q}$ 
({\sc YaJEM}) is physically successful compared to the BW model ({\sc YaJEM+BW}) 
at describing the CMS data and provides a mean $\hat{q}\sim2.4$ ${\rm GeV}^{2}/{\rm fm}$
with $\Delta Q^2\sim6$ ${\rm GeV}^{2}$ for a medium of length $L=2.5$ fm in the
hydrodynamical description of the QGP. The biased FFs and ratios are mainly dominated 
by quark-initiated showers and are well described by 
{\sc YaJEM} over the $\xi$ interval for all $P_T$ ranges of the CMS data. 
As a consequence of the jet quenching phenomena, a weak suppression of the hadron yield
in the intermediate region $0.4\lesssim z\lesssim0.7$ as well as the increase of soft 
hadrons in the softest region $0.7 \lesssim z\lesssim 0.015$ are well described by {\sc YaJEM}. 
The comparison of the biased versus unbiased FFs shows the importance of an accurate simulation of 
the jet-finding strategy, which suppresses the relevant physics of the 
jet quenching and therefore, information is lost concerning the early 
stage of jet evolution and its interaction with the QGP. Indeed, the trigger bias 
suppresses the range of possible medium modifications brought by the medium-induced 
soft gluon radiation \cite{Renk:2012ve}. Since these results are model-dependent, 
further comparison of the data with other energy loss event generators such as 
{\sc jewel}~\cite{Zapp:2008gi} and {\sc q-pythia}~\cite{Armesto:2009fj} are certainly interesting to 
unravel the details of parton energy loss in QCD matter.
 
\section*{Acknowledgements}

R.P.R. strongly thanks Beomsu Chang, David d'Enterria and Jiri Kral for useful 
discussions, comments on the manuscript and their expert help in numerical calculations. 

\bibliographystyle{h-elsevier3}
\bibliography{mybib}

\end{document}